\documentclass[12pt]{article}
\usepackage[dvips]{graphicx}
\usepackage{amssymb}
\usepackage{amsmath}
\usepackage{here}
\usepackage{color}
\usepackage{subfigure}

\newcommand{\bear}{\begin{array}}  
\newcommand {\eear}{\end{array}}
\newcommand{\bea}{\begin{eqnarray}}   
\newcommand{\eea}{\end{eqnarray}}
\newcommand{\beq}{\begin{equation}}   
\newcommand{\eeq}{\end{equation}}
\newcommand{\bef}{\begin{figure}}  \newcommand 
{\eef}{\end{figure}}
\newcommand{\bec}{\begin{center}}  \newcommand 
{\eec}{\end{center}}

\renewcommand{\thefootnote}{\fnsymbol{footnote}}
\begin{document}

\begin{titlepage}

\begin{flushright}
ICRR-Report-606-2011-23\\
\end{flushright}

\vskip 2.0cm

\begin{center}

{\large \bf Forecast constraints on cosmic string parameters from
    gravitational wave direct detection experiments }

\vskip 1.2cm

Sachiko Kuroyanagi$^a$,\footnote[1]{skuro@icrr.u-tokyo.ac.jp}
Koichi Miyamoto$^a$,
Toyokazu Sekiguchi$^{b}$,
Keitaro Takahashi$^{c}$\\
and Joseph Silk$^{d,e}$

\vskip 0.4cm

{\it $^a$Institute for Cosmic Ray Research,
University of Tokyo, Kashiwa 277-8582, Japan}\\
{\it $^b$Department of Physics and Astrophysics, Nagoya University, Nagoya 464-8602, Japan}\\
{ \it $^c$Faculty of Science, Kumamoto University, 2-39-1, Kurokami, Kumamoto 860-8555, Japan}\\
{\it $^d$Institut d' Astrophysique, 98bis Boulevard Arago, Paris 75014, France}\\
{\it $^e$Department of Physics, University of Oxford, Keble Road, Oxford, OX1 3RH, UK}

\vskip 1.2cm

\date{\today}

\begin{abstract} 
  Gravitational waves (GWs) are one of the key signatures of cosmic
  strings.  If GWs from cosmic strings are detected in future
  experiments, not only their existence can be confirmed but also
  their properties might be probed.  In this paper, we study the
  determination of cosmic string parameters through direct detection
  of GW signatures in future ground-based GW experiments.  We consider
  two types of GWs, bursts and the stochastic GW background, which
  provide us with different information about cosmic string
  properties.  Performing the Fisher matrix calculation on the cosmic
  string parameters, such as parameters governing the string tension
  $G\mu$ and initial loop size $\alpha$ and the reconnection
  probability $p$, we find that the two different types of GW can
  break degeneracies in some of these parameters and provide better
  constraints than those from each measurement.
\end{abstract}

\end{center}
\end{titlepage}

\renewcommand{\thefootnote}{\arabic{footnote}}

\section{Introduction}

Cosmic strings are one-dimensional topological defects which can be
formed at a phase transition in the early universe
\cite{Kibble:1976sj} (as a review, see \cite{Vilenkin}).  Strings in
superstring theory which have cosmological length, so-called cosmic
super strings, can also appear after the stringy model of inflation
and behave like cosmic strings
\cite{Sarangi:2002yt,Jones:2003da,Dvali:2003zj}.  They form the highly
complicated string network, including infinite strings, which stretch
across Hubble horizon, and closed loops, whose size is much smaller
than the Hubble scale, and leave cosmological effects in many ways
through their nonlinear evolution.  They therefore have attracted
strong attention since the possibility of their existence was pointed
out.  Considerable research concerning their observational signals and
their detectability have been done so far.  If their signals can be
observed precisely enough, not only the existence of cosmic strings
will be confirmed but their properties also might be studied.  Since
we can probe physics beyond the standard model of particle physics
such as grand unified theory or superstring theory through the
study of cosmic strings, it is very interesting and meaningful to
examine how the properties of cosmic strings can be constrained
through future experiments and observations.

Among important signals of cosmic strings are gravitational waves
(GWs)
\cite{Vilenkin:1981bx,Hogan:1984is,Caldwell:1991jj,Caldwell:1996en,Damour:2000wa,Damour:2001bk,Damour:2004kw,Siemens:2006yp,DePies:2007bm,Kawasaki:2010yi,Olmez:2010bi,Binetruy:2012ze}.
The main source of GWs in the string network is cusps on loops.
\footnote{GWs from kinks may dominate that from cusps in the case of
  cosmic superstring \cite{Binetruy:2010cc,Bohe:2011rk}. However their
  contribution strongly depends on the fraction of loops with
  junctions.  So we do not consider their contribution in this
  paper. }  A cusp is a highly Lorentz boosted region on a loop which
appears $\mathcal{O} (1)$ times in an oscillation period of the loop.
Beamed GW bursts are emitted from cusps and can be detected directly
as strong but infrequent bursts as well as in the form of a stochastic
GW background, which consists of many small bursts overlapping each
other \cite{Damour:2000wa,Damour:2001bk}.  The rate of GW bursts and
the spectrum of the GW background depend on the parameters which
characterize the string network.  Conversely, we can constrain the
parameters from the fact that GWs from cosmic strings have never been
detected, or even we can determine the values of them through future
observations if GWs from cosmic strings are detected.

So far, constraints on cosmic string parameters have been imposed by
LIGO from nondetection of either bursts or GW background
\cite{Abbott:2009rr,Abbott:2009ws}.  Currently LIGO is undergoing a
major upgrade (Advanced LIGO) \cite{Sigg:2008zz} and will increase the
detector sensitivity by more than an order of magnitude.  Furthermore,
several new ground-based detectors would be built along the same time
line as LIGO, such as KAGRA in Japan \cite{Kuroda:2010zzb} and
Advanced Virgo in Italy \cite{Accadia:2011zzc}.  These additional
detectors form a worldwide network of gravitational-wave
observatories, which can be a more powerful tool to search for
signatures of cosmic strings.
 
Cosmic strings can be characterized by the following parameters.  The
most important one is the tension $\mu$, the energy stored per unit
length in a cosmic string, which is often written in the form of its
product and Newton constant $G\mu$.  For field theoretic cosmic
strings, $\mu$ is roughly the square of the energy scale at the phase
transition which leads to the appearance of cosmic strings.  If they
appear at the spontaneous breaking of the symmetry of grand unified
theory, the expected value of $G\mu$ is of $\mathcal{O} (10^{-6})$ or
$\mathcal{O} (10^{-7})$.  For cosmic superstrings, $\mu$ is
proportional to the square of the string scale, but it also depends on
other parameters such as the warp factor of the extra dimension where
the cosmic superstring is located.  Therefore, the tension of cosmic
superstring can take a broad range of values.  Since $G\mu$ determines
not only the typical amplitude of GWs emitted from loops but also the
lifetime of loops, it affects the amplitude of GW bursts and the GW
background as well as the rate of bursts and the spectral shape of the
GW background.

The second one is the loop size $\alpha$.  It is well known that
infinite strings reach the scaling regime, where the curvature radius
and the interval of strings are comparable to the Hubble radius.  They
have to lose their length by releasing loops continuously in order to
maintain the scaling.  It is considered that the typical size of
initial loops also obey the scaling law and it is often written in the
form of $\alpha t$.  Although many analytic or numerical studies have
been conducted to determine the value of $\alpha$
\cite{Bennett:1987vf,Allen:1990tv,Vincent:1996rb,Vanchurin:2005pa,Ringeval:2005kr,Martins:2005es,Olum:2006ix,Siemens:2002dj,Polchinski:2006ee,Dubath:2007mf,Vanchurin:2007ee,Lorenz:2010sm},
it still remains controversial.
We treat it as a free parameter in this paper.  This parameter
$\alpha$ also affects the rate of GW bursts and the spectrum of the GW
background, since lifetime of loops and frequency of GWs emitted from
loops depend on the loop size.

The third one is the reconnection probability $p$.  It is well known
that, in almost all models, field theoretic cosmic strings necessarily
reconnect when they collide.  On the other hand, cosmic superstrings
can have the reconnection probability much smaller than $1$, because a
collision between them is a quantum process and they can miss each
other in the extra dimension
\cite{Jackson:2004zg,Hanany:2005bc,Jackson:2007hn}.  The reduced
reconnection probability leads to an inefficient loss of length of
infinite strings and eventually causes an enhancement of their density
\cite{Sakellariadou:2004wq,Avgoustidis:2005nv}.  Since the number of
loops accordingly increases, the amplitude of GW background and the
burst rate are enhanced as $p$ decreases.

In this paper, we calculate the spectrum of the GW background and the
rate of "rare bursts", which are isolated and whose amplitudes exceed
that of the background, and study their dependencies on above three
parameters.  Then we find the parameter region which is excluded by
current experiments or can be searched by future experiments.
Finally, assuming that GWs will be detected by ground-based GW
detectors, we use the Fisher matrix formalism in order to investigate
the degree to which the cosmic string parameters are constrained by
future experiments.  It is notable that the rare bursts and the
stochastic GW background have different information and constraints
from them break the parameter degeneracies each other.

This paper is constructed as follows.  In the next section, we show
the model of the string network assumed in this paper.  In Sec. III,
we describe the formalism for calculation of the burst rate and the GW
background spectrum and show some examples assuming specific parameter
values.  In Sec. IV, we find the parameter region which can be
probed by upcoming GW experiments such as Advanced LIGO and the world
wide GW network.  Furthermore, we calculate the Fisher information
matrix and predict the combined constraints from burst detection and
GW background measurements, choosing parameter values where both are
accessible by future experiments.  The last section is devoted to the
summary.

\section{Analytic model of cosmic string network}

\subsection{Infinite strings}

\begin{figure}[tbp]
\begin{center}
\includegraphics[width=80mm]
{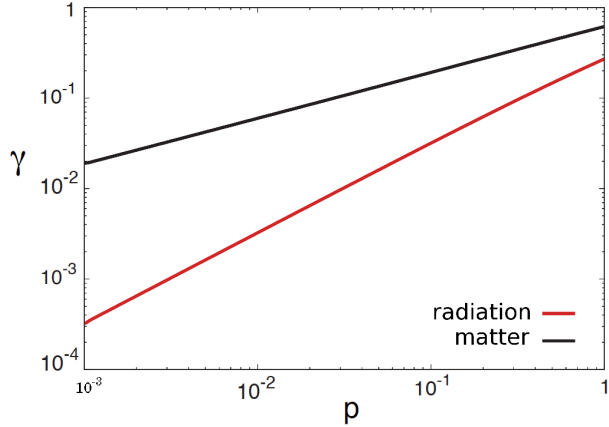}
\caption{The values of $\gamma$ plotted as a function of $p$ 
for the radiation and matter-dominated era.}
\label{fig:gamma}
\end{center}
\end{figure}

We adopt the model in \cite{Avgoustidis:2005nv,Takahashi:2008ui},
which is based on the velocity-dependent one-scale model
\cite{Martins:1996jp}.  The network of infinite strings can be
considered as a random walk, so their total length $L$ in volume $V$
can be written as
\beq
L=\frac{V}{\xi^2},
\label{lengthinV}
\eeq
where $\xi$ is the correlation length, which corresponds to the
typical curvature radius and interval of infinite strings.  The
equations for $\gamma\equiv\xi/t$ and the root mean velocity of
infinite strings $v$ are %
given by
\beq
\frac{t}{\gamma}\frac{d\gamma}{dt}=-1+\nu +\frac{\tilde{c}pv}{2\gamma}+\nu v^2,
\label{gammaeq}
\eeq 
\beq
\frac{dv}{dt}= (1-v^2)H (\frac{k (v)}{\nu\gamma}-2v),
\label{veq}
\eeq
where $k(v)=\frac{2\sqrt{2}}{\pi}\frac{1-8v^6}{1+8v^6}$
\cite{Martins:2000cs}, $H$ is the Hubble parameter, and the scale
factor $a$ is parametrized as $a (t)\propto t^{\nu}$.  The constant
parameter $\tilde{c}$ represents the efficiency of loop formation and
we set it to be $\tilde{c}=0.23$ according to
Ref. \cite{Takahashi:2008ui}.  For the cosmic string whose
reconnection probability is $p<1$, the probability of loop formation
when a string self-reconnects is also $p$, then $\tilde{c}$ is
replaced with $\tilde{c}p$.  In the scaling regime, $\gamma$ and $v$
become constant.  We can get their asymptotic values by setting
$d\gamma/dt$ and $dv/dt$ to be $0$. Figure $\ref{fig:gamma}$ shows the
value of $\gamma$ in the radiation-dominated era ($\nu=1/2$) and the
matter-dominated era ($\nu=2/3$) as a function of the reconnection
probability $p$.  This figure shows that $\gamma$ is proportional to
$p$ and $p^{1/2}$ in the radiation and matter-dominated era,
respectively.  Hereafter, we denote the values of $\gamma$ in the
radiation and matter-dominated era as $\gamma_r$ and $\gamma_m$.
Since the effect of the time dependence of $\gamma$ around the
matter-radiation equality is small for the parameters we set below, we
approximate $\gamma$ as a step function,
\beq \gamma (z)=
\begin{cases}
\gamma_r & ;z>z_{eq} \\
\gamma_m& ;z<z_{eq}
\end{cases}
,
\eeq
where $z_{eq}$ is the redshift at the matter-radiation equality.

\subsection{Loops}
In order to maintain the scaling, infinite strings have to
continuously release their length in the form of loops.  The released
length in a Hubble volume per Hubble time is comparable to the length
of infinite strings in a Hubble volume.  Since the length of a loop
formed at time $t$ is given by $\alpha t$, the number density of loops
produced between time $t$ and $t+dt$ is
\beq
\frac{dn}{dt}(t)dt=\frac{dt}{\alpha \gamma^2 t^4}.
\eeq
The number density of loops is diluted proportional to $a^{-3}$ by
cosmic expansion.  Therefore, the number density of loops formed
between $t_i$ and $t_i+dt_i$ at time $t$ is
\beq
\frac{dn}{dt_i} (t,t_i)dt_i=\frac{dt_i}{\alpha \gamma^2 t_i^4}\left (\frac{a (t_i)}{a (t)}\right)^3.
\eeq
After loop formation, it continues to shrink by emitting its energy as
GWs.  The energy spectrum of GWs from a loop of circumference $l$ per
unit time is given by
\beq
\frac{d\dot{E}}{df}\sim G\mu^2l^{-1/3}f^{-4/3} \label{dEdf}
\eeq
with a low frequency cutoff at $f\sim l^{-1}$.  The total energy
emission rate is
\beq
\dot{E} = \Gamma G\mu^2,
\eeq
where $\Gamma$ is set to be $50$ in this paper.  Then, the length of a
loop formed at $t_i$ is
\beq
l (t,t_i)=\alpha t_i - \Gamma G\mu  (t-t_i), \label{length}
\eeq
at time $t$.  Lifetime of a loop formed at $t_i$ is given by
$\frac{\alpha}{\Gamma G\mu}t_i$.  If $\alpha < \Gamma G\mu$, they are
short-lived, that is, they decay within a Hubble time.  If $\alpha >
\Gamma G\mu$, loops are long-lived, that is, they live longer than a
Hubble time.  In this case, loops have a wide range of length from
$\alpha t$ to $\Gamma G\mu t$, which corresponds to loops just formed
and expiring at some time.  The most numerous loops are those of
length comparable to $\sim \Gamma G\mu t$.

\section{Gravitational waves from cosmic string loops}
\label{secGW}
In this section, we describe the calculation procedure for the burst
rate and the spectrum of the GW background, following the formalism in
Ref. \cite{Damour:2001bk}.  Using the result, we show parameter
dependence of the burst rate and the background spectrum for several
parameter sets.

\subsection{Formalism}
The linearly polarized waveform of a GW burst 
emitted in a direction $\mathbf{n}$ 
by a loop with circumference $l$ at redshift $z$ is expressed
as
\beq
h_{\mu\nu} (t,\mathbf{n})=\int df h (f,z,l) e^{-2\pi i ft}e^+_{\mu\nu} (\mathbf{n}) \times \Theta (\mathbf{n}\cdot \mathbf{n}_c - \cos\theta_m (f,z,l)) \times \Theta (1-\theta_m (f,z,l)), \label{waveform}
\eeq
where $\mathbf{n}_c$ is the direction of the center of the burst,
which coincides with that of the derivative of the right or left
moving mode of the loop, and $\theta_m$ is the beaming angle of the GW
burst which is given by
\beq
\theta_m (f,z,l)= ( (1+z)fl)^{-1/3}. \label{thetam}
\eeq
The first Heaviside step function $\Theta$ in Eq.  (\ref{waveform})
reflects the fact that the burst is emitted into a limited angle, and
the second one means that it has a low frequency cutoff at $f\lesssim
l^{-1}$ when it is emitted.  The polarization tensor (for plus
polarization) is expressed as
$e^+_{\mu\nu}=l_{\mu}m_{\nu}-l_{\nu}m_{\mu}$, where $l_{\mu}=
(0,\mathbf{l}), m_{\mu}= (0,\mathbf{m})$ and $\mathbf{l}$ and
$\mathbf{m}$ are unit vectors orthogonal to $\mathbf{n}$ and each
other.

A GW burst from a loop is most efficiently generated at frequency
comparable to $l^{-1}$ and the amplitude of higher frequency modes
decrease in proportion to $f^{-4/3}$ \footnote{ Strictly speaking, Eq.
  (\ref{strain}) is valid only for modes whose frequency is much
  higher than $\sim l^{-1}$ when the GW is emitted and the amplitude
  of low frequency modes, $f\sim l^{-1}$, depends on the detail of the
  oscillation of the loop.  However, it is known from numerical
  studies that it is a good approximation to apply the power law as in
  Eq. (\ref{strain}) to low frequency modes for small loops
  \cite{Allen:1991bk}.  }. The Fourier transform of the GW amplitude
is given by \footnote{ Note that this definition $h$ is the same as
  Ref. \cite{Siemens:2006yp} and different from
  Ref. \cite{Damour:2001bk} by a factor of $f$.  }
\beq
h (f,z,l)\sim\frac{4\pi  (12)^{4/3}}{ (2\pi)^{1/3} (3\Gamma (1/3))^2}\frac{G\mu l}{ ( (1+z)fl)^{1/3}r (z)f}\simeq 2.68\frac{G\mu l}{ ( (1+z)fl)^{1/3}r (z)f}, \label{strain}
\eeq
where
\beq
r (z)=\int^z_0 dz^{\prime}\frac{1}{H (z^{\prime})},
\eeq
and
$H(z)=H_0\left(\Omega_{\Lambda}+\Omega_m(1+z)^3+\Omega_r(1+z)^4\right)^{1/2}$
is the Hubble parameter at redshift $z$, where $H_0$ is the Hubble
parameter of today and $\Omega_{\Lambda}$, $\Omega_m$, and $\Omega_r$
are the present values of the ratio of the energy density to the total
energy density for dark energy, matter and radiation, respectively.
Note that one can confirm that this frequency leads to the energy
spectrum of GWs from a loop given in Eq. (\ref{dEdf}) (see, for
example, Ref. \cite{CandG}).

The number of GWs coming to the Earth per unit time, emitted at
redshift between $z$ and $z+dz$ by loops formed between $t_i$ and
$t_i+dt_i$ is
\beq
\frac{dR}{dzdt_i}dzdt_i=\frac{1}{4}\theta_m (f,z,l)^2\frac{2c}{ (1+z)l (t (z),t_i)}\frac{dn}{dt_i} (t (z),t_i)dt_i\frac{dV}{dz}dz\times\Theta (1-\theta_m (f,z,l)). \label{dR_dtdti}
\eeq
Here, $\frac{1}{4}\theta_m (f,z,l)^2$ is the factor which
represents the fraction of GW bursts beamed towards the Earth.  Cusp
formation is expected to occur $\mathcal{O} (1)$ times in an
oscillation period, which is characterized by parameter $c$.  We set
it to be $1$ in this paper.  The factor $\frac{dV}{dz}dz$ is the
volume between $z$ and $z+dz$ at time $t (z)$ and given by
\beq
\frac{dV}{dz} (z)=\frac{4\pi a^2 (z)r^2 (z)}{H (z) (1+z)}.
\eeq

Since the observables are the amplitude and the frequency of GW
bursts, we need a prediction of the burst rate expressed in terms of
the given amplitude and frequency.  Using Eqs. (\ref{length}) and
(\ref{strain}), we can rewrite Eq. (\ref{dR_dtdti}) to express the
number of GWs coming per unit time which were emitted at redshift $z$
and which have frequency $f$ and amplitude $h$ at the present time
\beq
\frac{dR}{dzdh} (f,h,z)=\frac{3}{4}\theta^2_m (f,z,l)\frac{c}{ (1+z)h}\frac{1}{\gamma^2 \alpha t^4_i}\frac{1}{\alpha+\Gamma G\mu}\left (\frac{a (t_i)}{a (t)}\right)^3 \frac{dV}{dz}\Theta  (1-\theta_m (f,z,l)),
\label{dR_dzdh}
\eeq
where $l$ and $t_i$ can be given as functions of $h$, $z$, and $f$,
\beq
l (f,h,z)=\left (\frac{hr (z)}{2.68G\mu} (1+z)^{1/3}f^{4/3}\right)^{3/2}, \label{l_of_hzf}
\eeq
\beq
t_i (f,h,z)=\frac{l (f,h,z)+\Gamma G\mu t (z)}{\alpha+\Gamma G\mu}. \label{t_i}
\eeq
By integrating Eq. (\ref{dR_dzdh}) in terms of $z$, we get the rate of
GWs for the given frequency and amplitude,
\beq
\frac{dR}{dh}=\int^{\infty}_0 dz \frac{dR}{dhdz}.
\eeq
Bursts overlapping each other form a stochastic background.  We adopt
the criterion in Ref. \cite{Siemens:2006yp}, which counts bursts
coming to the Earth with a time interval shorter than the oscillation
period of themselves as a component of the GW background.  Such bursts
have amplitude smaller than $h_*$, which is determined for a given
frequency as
\beq
\int^{\infty}_{h_*} dh\frac{dR}{dh}=\int^{\infty}_{h_*}dh \int^{\infty}_0dz \frac{dR}{dzdh}=f.
\eeq
The amplitude of a stochastic GW background is commonly expressed by
$\Omega_{\rm GW} (f)\equiv (d\rho_{\rm GW}/d\ln f)/\rho_{cr}$ where
$\rho_{\rm GW}$ is the energy density of the GWs and $\rho_{cr}$ is
the critical density of the Universe. Then the spectral amplitude of
the GW background is given by summing up all contributions from small
bursts,
\beq
\Omega_{\rm GW} (f)=\frac{2\pi^2}{3H_0^2}f^3\int^{h_*}_0dhh^2\frac{dR}{dh}.
\label{OmegaGW}
\eeq

\subsection{Parameter dependence of the burst rate and the GW
  background spectrum} \label{Param_dep_spec}

\begin{figure}[!t]
\begin{center}
\includegraphics[width=110mm]
{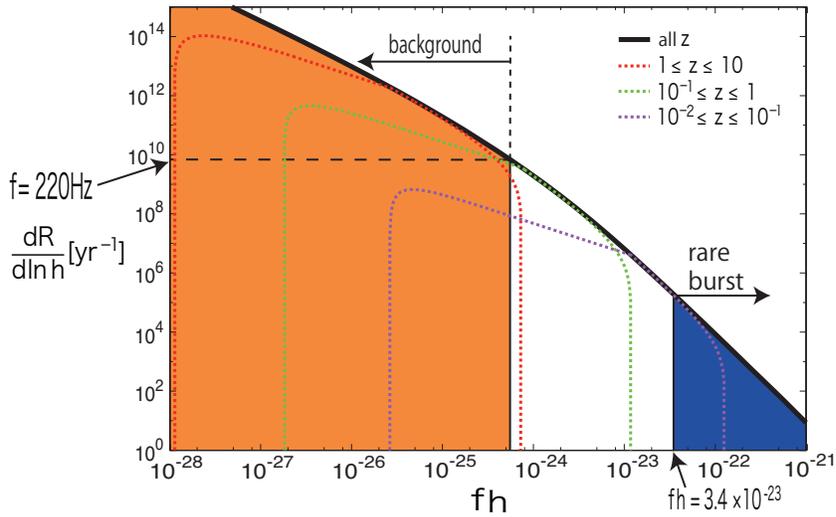}
\caption{ The rate of GW burst $dR/d{\ln}h$ as a function of $fh$, for
  $G\mu=10^{-7}$, $\alpha=10^{-16}$, $p=1$, and $f=220{\rm Hz}$,
  represented by the black thick line.  Bursts in the left region
  (orange) form the GW background and those in the right region (blue)
  are observed as an isolated burst. The red, green and purple dotted
  line represent the contribution from $1\le z \le 10$, $0.1 \le z \le
  1$, and $10^{-2} \le z \le 0.1$, respectively.  }
\label{fig:dRdlnh_ex}
\end{center}
\end{figure}

\begin{figure}[p]
\begin{minipage}{1\hsize}
\begin{center}
\subfigure[For $G\mu=10^{-7}$, $p=1$, $f=220{\rm Hz}$, and various $\alpha$.]{
\includegraphics[width=90mm]
{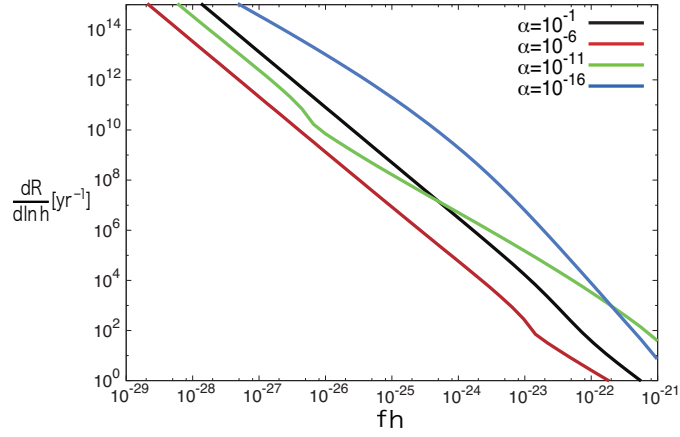}
\label{fig:dRdlnh_alpha}
}
\subfigure[For $\alpha=10^{-16}$, $p=1$, $f=220{\rm Hz}$, and various $G\mu$.]{
\includegraphics[width=90mm]
{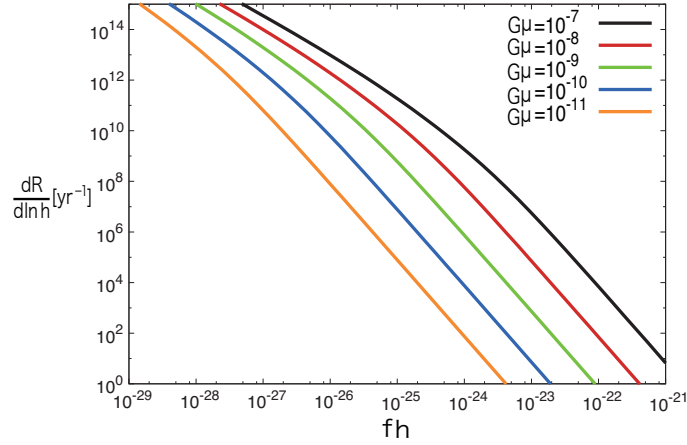}
\label{fig:dRdlnh_Gmu}
}
\subfigure[For $G\mu=10^{-7}$, $\alpha=10^{-16}$, $f=220{\rm Hz}$, and various $p$.]{
\includegraphics[width=90mm]
{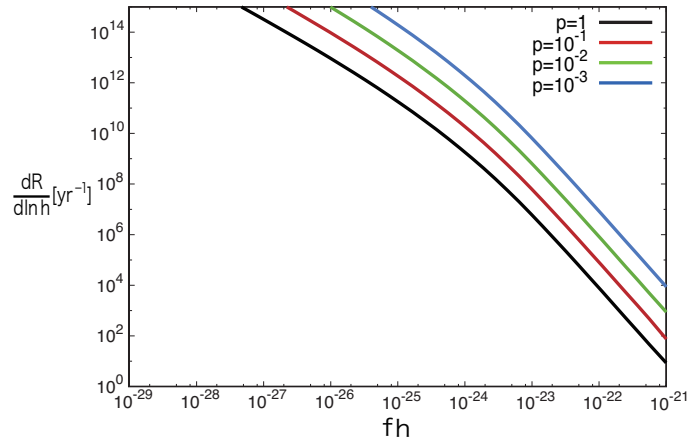}
\label{fig:dRdlnh_p}
}
\end{center}
\end{minipage}
\caption{The burst rate $dR/d\ln h$ in terms of $fh$ for various
  parameter sets.}
\end{figure}

\begin{figure}[t]
\begin{tabular}{cc}
\begin{minipage}{0.5\hsize}
\begin{center}
\subfigure[For $G\mu=10^{-7}$, $p=1$, and various $\alpha$.]{
\includegraphics[width=80mm]
{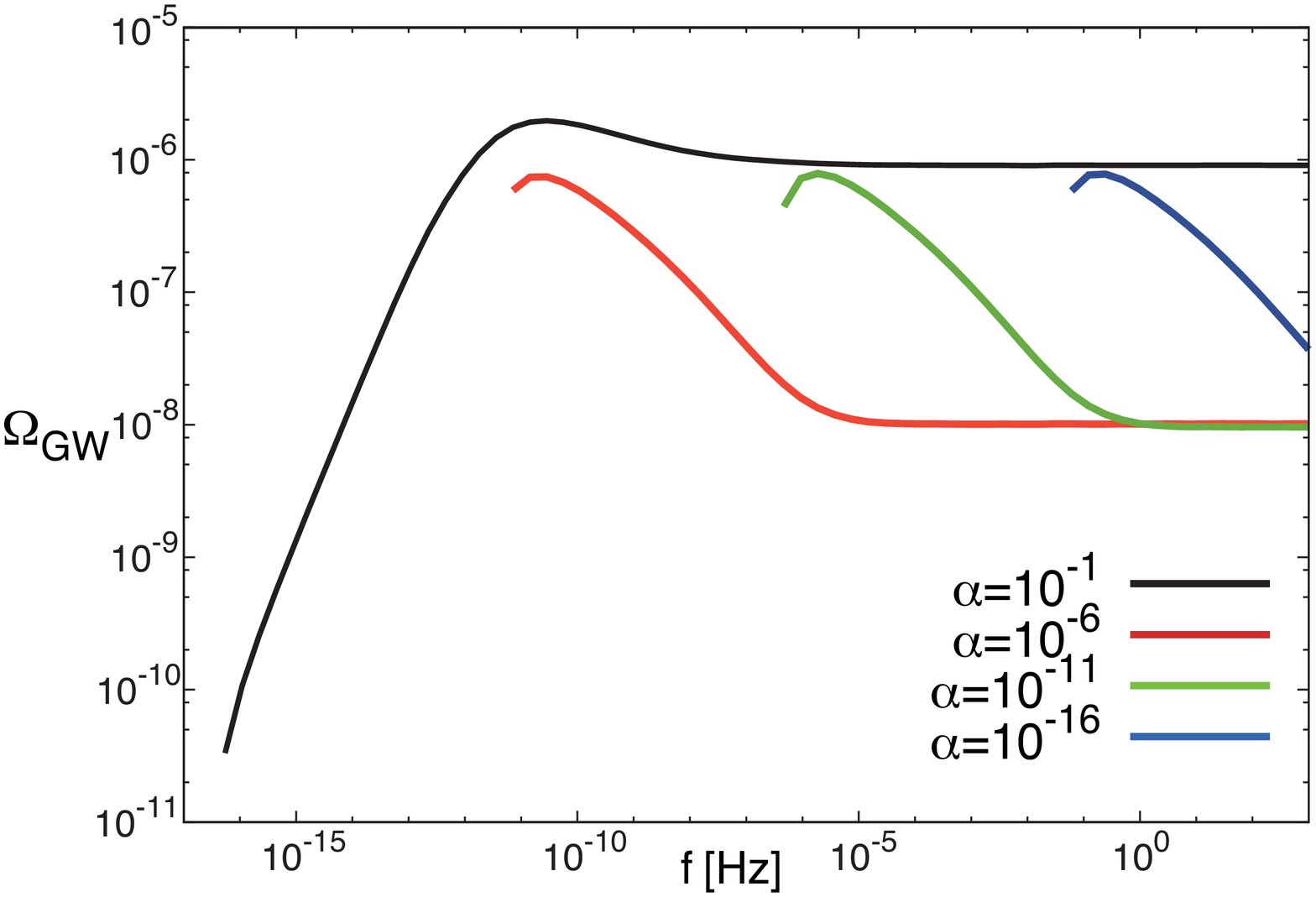}
\label{fig:Omegagw_alpha}
}
\end{center}
\end{minipage}

\begin{minipage}{0.5\hsize}
\begin{center}
\subfigure[For $\alpha=10^{-16}$, $p=1$, and various $G\mu$.]{
\includegraphics[width=80mm]
{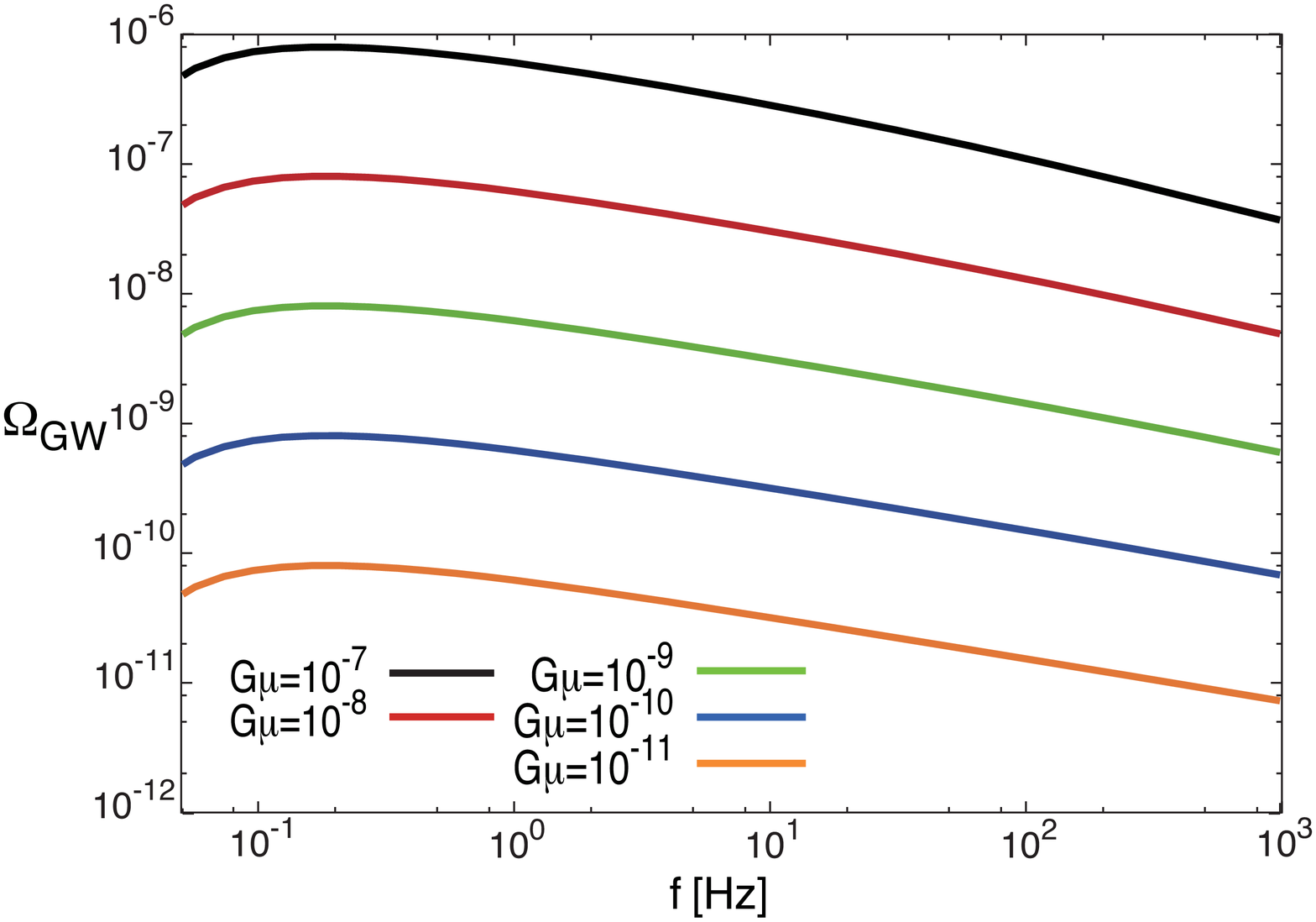}
\label{fig:Omegagw_Gmu}
}
\end{center}
\end{minipage}
\end{tabular}
\caption{The background spectrum $\Omega_{\rm GW}$ versus present
  frequency $f$ for various parameter sets.}
\end{figure}

Here, we show how the burst rate $dR/dh$ as a function of amplitude
$h$ depends on parameters $G\mu$, $\alpha$, and $p$.  In Fig.
\ref{fig:dRdlnh_ex}, we plot $dR/d{\ln} h$ versus $h$ for
$G\mu=10^{-7}$, $\alpha=10^{-16}$, $p=1$, which is evaluated at
$f=220{\rm Hz}$, the best-sensitivity frequency of Advanced LIGO.

We set cosmological parameters, to be the WMAP 7-year mean values
\cite{Komatsu:2010fb}: the dark energy density divided by the critical
density $\Omega_{\Lambda}=0.728$, the dark matter density divided by
the critical density $\Omega_{m}=0.272$, the current CMB temperature
$T_0=2.725{\rm K}$, and the Hubble constant $H_0=70.4 {\rm km}/{\rm
  s}/{\rm Mpc}$.  We see the natural tendency that stronger bursts
have a lower rate.  In this figure, we also show the ranges of $h$
which correspond to bursts observed as rare bursts or the GW
background by Advanced LIGO.  GW bursts whose amplitude are larger
than the detector sensitivity, which corresponds to $fh=3.4\times
10^{-23}$ at $f=220{\rm Hz}$ for Advanced LIGO, are observed as rare
bursts.  In contrast, GW bursts whose rate is larger than their
frequency, $f\sim 220{\rm Hz}$ at Advanced LIGO, are measured as the
GW background. \footnote{Note that bursts in the middle amplitude
  range between the rare burst and GW background regions may be
  detected as unresolved sources in the GW background
  \cite{Regimbau:2011bm} .  Non-Gaussian measurements of the GW
  background may be useful to characterize their contributions
  \cite{gr-qc/0210032,Seto:2008xr,Seto:2009ju}.}
    
GW bursts in each amplitude bin consist of bursts from different
redshifts.  In order to illustrate which redshift mainly contributes
to bursts at a given amplitude, we also plot the contributions to
$dR/d\ln h$ from different redshift ranges in Fig.
\ref{fig:dRdlnh_ex}.  The red, green, and purple dotted line
correspond to bursts from $1 \le z \le 10$, $10^{-1} \le z \le 1$, and
$10^{-2} \le z \le 10^{-1}$, respectively.  This indicates that, in
this parameter set, bursts detectable by Advanced LIGO come from
redshift lower than $z\sim 10^{-2}$, and the GW background consists of
bursts emitted at redshifts higher than $z\sim 1$.

Figures \ref{fig:dRdlnh_alpha}, \ref{fig:dRdlnh_Gmu}, and
\ref{fig:dRdlnh_p} show dependencies of the burst rate on parameters,
$G\mu$, $\alpha$, and $p$, respectively.  The parametric dependence of
the burst rate is the key in studying constraints on cosmic string
parameters by GW direct detection experiments, since it determines the
direction of the parameter degeneracy, The details are investigated in
Appendix \ref{app1}.  Here we only give some short explanations.  The
dependence on $p$ is simplest because a small value of $p$ simply
enhances the rate through the factor $\gamma^{-2}$ in Eq.
(\ref{dR_dzdh}).  A large value of $G\mu$ basically leads to a larger
burst rate because of the enhancement of the amplitude of each burst.
However the actual dependence is more nontrivial since the variation
of $G\mu$ also changes the typical lifetime of loops, which leads to
difference in their density.  Variation of $\alpha$ also affects the
lifetime of loops, as well as the initial number of loops.  Therefore
the dependence on $\alpha$ is also not easy to simplify.

We also show the spectrum of the GW background, $\Omega_{\rm GW}$, for
different parameter sets in Fig. \ref{fig:Omegagw_alpha} and
Fig. \ref{fig:Omegagw_Gmu}.  We also explain details of the parameter
dependence of $\Omega_{\rm GW}$ in Appendix \ref{app2}.  Physically,
$\Omega_{\rm GW}$ at each frequency reflects the energy ratio of
expiring loops to the total energy of the Universe at the redshift
corresponding to $\min\{\alpha,\Gamma G\mu\}\times t\sim f^{-1}
(1+z)^{-1}$ for the radiation-dominated universe, or, for low
frequency modes, it reflects the energy of GWs emitted recently.

For these reasons, rare bursts and the background spectrum carry
information of cosmic strings at different redshifts, and have
different dependence on cosmic string parameters.  This is one of the
reason they provide different directions of parameter degeneracy in
parameter constraints by burst detection and the GW background
measurements, as we shall present in the next section.

\section{Constraints on cosmic string parameters from future GW
  experiments}
In this section, we investigate how accurately the cosmic string
parameters can be determined if GWs from cosmic strings are detected
by future experiments.  We consider the case where both the GW bursts
and the GW background are detected and show their constraints
complement each other.  Before that, we compute the sensitivities of
GW detectors and show the accessible parameter space by current and
future experiments.

\subsection{Sensitivity for GW detection}
In this paper, we assume the worldwide GW network, consists of
Advanced LIGO pairs, KAGRA, VIRGO.  Additional detectors
improve angular resolution and enable precise determination of
intrinsic amplitude and polarization of the incoming GWs.  Also, since
these worldwide GW detectors are directed to different regions of the
sky, they provide almost all sky coverage and increase the number of
detection events, while each GW detector has limited sky coverage.
The detector network also improves sensitivity of the GW background
search with the cross-correlation analysis.
  
\subsubsection{Burst}
\label{burst_detectability}
As shown in Eq.  (\ref{strain}), the burst signal from a cosmic string
cusp would be linearly polarized and have the frequency dependence of
$f^{-4/3}$ with cutoffs at low and high frequencies.  We assume that
the spectrum has the form of
\begin{equation}
h^+ (f)=Af^{-4/3}\Theta (f_h-f)\Theta (f-f_l),
\end{equation}
where the amplitude $A$ can be read from Eq.  (\ref{strain}).  The low
frequency cutoff corresponds to the size of the loop, whose scale is
typically cosmological.  So, usually, the cutoff frequency is much
lower than the lower frequency limit of direct detection experiments.
The high frequency cutoff depends on the viewing angle as $f_h\sim 2/
(\theta_{\rm obs}^3L)$, where $\theta_{\rm obs}$ is defined as the
angular separation between the observer's line of sight and the beam
direction of the GW emission.

The total output of the detector is written as a combination of GW
signal $h (t)=F^+h^+ (t)$ and detector noise $n (t)$, $s (t)=h (t)+n
(t)$, where $F^+$ describes the detector response to plus polarized
GWs.  This function depends on the sky location of the GW source, the
polarization angle and the configuration of the detectors, and can be
interpreted as the sky area covered by the experiments.  For a single
detector, one may use the all sky-averaged value for orthogonal arm
detectors, $\overline{F^+}\sim 1/\sqrt{5}$.  For next generation
detector network (Advanced LIGO, KAGRA and VIRGO), we assume
that the GW detector network has $100\%$ visibility over the whole sky
so that $\overline{F^+}\sim 1$.

The search for GW bursts signals from cosmic strings is usually
performed via matched filtering \cite{Siemens:2006vk,Abbott:2009rr}.
The template for cosmic string bursts is usually taken as
\begin{equation}
\tau (f)=f^{-4/3}\Theta (f_h-f) (f-f_l).
\end{equation}
The template is normalized by dividing by $\sigma=\sqrt{ (\tau|\tau)}$,
where the inner product is defined as
\begin{equation}
 (x|y)\equiv 4 \Re\int^{\infty}_{0}df\frac{x (f)y^* (f)}{S_n (f)}.
\end{equation}
Thus the normalized template is $\hat{\tau}\equiv\tau/\sigma$, which
satisfies $(\hat{\tau}|\hat{\tau})=1$.  The noise spectral density
$S_n (f)$ is defined by $\langle n (f)^*n (f)\rangle\equiv S_n
(f)\delta (f-f^\prime)/2$, where $\langle\cdots\rangle$ denotes
the ensemble average.  The noise spectral density for
Advanced LIGO is given by \cite{Key:2008tt}
\begin{equation}
S_n (f)=10^{-49}\left[x^{-4.14}-\frac{5}{x^2}+111\left (\frac{2-2x^2+x^4}{2+x^2}\right)\right]{\rm Hz}^{-1},
\label{Adv-LIGO}
\end{equation}
where $x=f/ (215{\rm Hz})$, and the noise for current LIGO is given by
\cite{Siemens:2006vk}.
\begin{eqnarray}
S_n (f)=
1.09\times 10^{-41}\left (\frac{30{\rm Hz}}{f}\right)^{28}+1.44\times 10^{-45}\left (\frac{100{\rm Hz}}{f}\right)^4\nonumber\\
+1.28\times 10^{-46}\left (1+\left (\frac{90{\rm Hz}}{f}\right)^{-2} \right)
{\rm Hz}^{-1}.
\label{LIGO}
\end{eqnarray}
For other GW network observatories, we assume all detectors have the
same sensitivity as Advanced LIGO.

The signal to noise ratio $\rho$ is given by the product of the signal
and the normalized template as $\rho\equiv (s|\hat{\tau})$.  This is
equivalent to calculate
\begin{equation}
\rho=\left[4\int^{f_h}_{f_l}df\frac{|h (f)|^2}{S_n (f)}\right]^{1/2}.
\end{equation}
The low frequency cutoff is determined by the detector's limitation,
which is taken to be $f_l=10$Hz for Advanced LIGO and $f_l=40$Hz for
current LIGO.  The high frequency cutoff is different for each burst,
depending on the viewing angle.  However, in most cases of detection,
it is distributed around the most sensitive frequency of the detector
\cite{Cohen:2010xd}.  Here, we take $f_h=220$Hz for Advanced LIGO and
$f_h=150$Hz for current LIGO.  Taking the detection threshold as
$\rho>4$ \cite{Damour:2000wa}, we find the future GW detector network
(Advanced LIGO sensitivity with full sky coverage $\overline{F^+}\sim
1$) can detect GW bursts whose amplitudes are larger than $A\simeq
2.1\times 10^{-22}$ ${\rm s}^{-1/3}$, which corresponds to $fh\simeq
3.4\times 10^{-23}$ at $f=220$Hz.  For current LIGO (LIGO sensitivity
with $\overline{F^+}\sim 1/\sqrt{5}$), the detection limit is $A\simeq
9.1\times 10^{-21}$ ${\rm s}^{-1/3}$ which corresponds to $fh\simeq
1.7\times 10^{-21}$ at $f=150$Hz.

\subsubsection{Stochastic GW background}
\label{GWBdetection}
The GW background is searched by correlating output signals of two or
multiple detectors.  One may define the cross correlation signal
between two detectors labeled by $I$ and $J$ as \cite{Allen:1997ad}
\begin{equation}
S=\int^{T/2}_{-T/2}dt\int^{T/2}_{-T/2}dt^{\prime} s_I (t)s_J (t^{\prime})Q (t,t^{\prime}),
\end{equation}
where $T$ is the observation time and $Q (t,t^{\prime})$ is a filter
function.  Using the fact that noises of different detectors have no
correlation each other, $\langle s_I (t)s_J (t^{\prime})\rangle\simeq
\langle h_I (t)h_J (t^{\prime})\rangle$, and transforming to Fourier
space, the mean value of the signal can be expressed as
\begin{equation}
\mu\equiv\langle S \rangle =\int^{\infty}_{-\infty}df\int^{\infty}_{-\infty}df^{\prime}\delta_T (f-f^{\prime})\langle\tilde{h}^*_I (f)\tilde{h}_J (f^{\prime})\rangle\tilde{Q} (f^{\prime}),
\end{equation}
where the tilde denotes Fourier-transformed quantities and $\delta_T
(f-f^{\prime})\equiv\int^{T/2}_{-T/2}dt e^{-2\pi ift}=\sin (\pi
fT)/\pi f$.  The response of the detector is given using $F^\lambda$
as
\begin{equation}
\tilde{h}_I (f)=\sum_\lambda\int d\hat{\bf \Omega} \tilde{h}_\lambda (f,{\bf \Omega})
e^{-2\pi if\hat{\bf \Omega}\cdot{\bf x}_I} F^\lambda_I (f,{\bf \Omega}),
\end{equation}
where $\lambda$ runs for both plus ($+$) and cross ($\times$)
polarization and ${\bf x}_I$ denotes the position of the detector.
Using the relation between the Fourier amplitudes $h_\lambda (f,{\bf
  \Omega})$ and $\Omega_{\rm GW}$,
\begin{equation}
\langle \tilde{h}^*_\lambda (f,{\bf \Omega})
\tilde{h}_{\lambda^{\prime}} (f^{\prime},{\bf \Omega^{\prime}})\rangle
=\frac{3H_0^2}{32\pi^3}\delta^2 ({\bf \Omega,\Omega^{\prime}})
\frac{1}{2}\delta_{\lambda\lambda^{\prime}}\delta (f-f^{\prime})
|f|^{-3}\Omega_{\rm GW} (|f|),
\end{equation}
the cross correlation signal is given by
\begin{equation}
\mu=\frac{3H_0^2}{20\pi^2} T
\int^{\infty}_{-\infty}df|f|^{-3} \gamma_{IJ} (f)
\Omega_{\rm GW} (f)\tilde{Q} (f),
\label{Eqmu}
\end{equation}
where we define the overlap reduction function as
\begin{equation}
\gamma_{IJ} (f)\equiv\frac{5}{8\pi}\int d\hat{\bf \Omega}
 (F^+_IF^+_J+F^{\times}_IF^{\times}_J)
e^{-2\pi if\hat{\bf \Omega}\cdot ({\bf x}_I-{\bf x}_J)}.
\end{equation}
We calculate the overlap reduction function following the procedure
given in Ref. \cite{Nishizawa:2009bf}, whose Table 2 or Table 3 provides the
relative positions of future ground-based GW detectors.  In the
weak-signal assumption, the variance of the correlation signal is
\begin{eqnarray}
\sigma^2&\equiv&\langle S^2\rangle-\langle S \rangle^2\approx\langle S^2\rangle
\\
&=&\int^{T/2}_{-T/2}dt\int^{T/2}_{-T/2}dt^{\prime} \langle s_I (t)s_J (t)s_I (t^{\prime})s_J (t^{\prime})\rangle Q (t) Q (t^{\prime}),
\end{eqnarray}
Using $\langle s_I (t)s_J (t)s_I (t^{\prime})s_J (t^{\prime})\rangle\simeq
\langle n_I (t)n_I (t^{\prime})\rangle\langle
n_J (t)n_J (t^{\prime})\rangle$ and transforming to Fourier space, this
can be expressed in terms of the noise spectral density $S_n (f)$ as
\begin{equation}
\sigma^2\approx\frac{T}{4}
\int^{\infty}_{-\infty}df S_{n,I} (|f|)S_{n,J} (|f|)|\tilde{Q} (f)|^2.
\label{Eqsigma}
\end{equation}

The signal to noise ratio is defined by $\rho\equiv\mu/\sigma$.
Choosing the optimal function to maximize the signal to noise ratio,
which is $\tilde{Q} (f)\propto \frac{\gamma_{IJ} (|f|)\Omega_{\rm
  GW} (|f|)}{|f|^3S_{n,I} (|f|)S_{n,J} (|f|)}$, we obtain
\begin{equation}
\rho_{IJ}=\frac{3H_0^2}{10\pi^2} \sqrt{2T}
\left[\int^{\infty}_{0}df\frac{|\gamma_{IJ} (f)|^2
\Omega_{\rm GW} (f)^2}{f^6S_{n,I} (f)S_{n,J} (f)}\right]^{1/2}.
\label{rhoIJ}
\end{equation}
For a network of $N$ detectors, we can make $N (N-1)/2$ independent
correlation signals.  One can calculate signal to noise ratio as
 (see Sec. V-C of Ref.  \cite{Allen:1997ad})
\begin{equation}
\rho=
\left[\sum^N_{I=1}\sum^N_{J<I}\rho_{IJ}^2\right]^{1/2}.
\end{equation}

Advanced LIGO detector pair would be able to detect the GW background
with $\rho>4$ if $\Omega_{\rm GW}>5.1\times 10^{-9}$ (for a flat
spectrum) with 3-year observation.  The multiple detector network
would reach $\Omega_{\rm GW}\sim 3.6\times 10^{-9}$, where we take the
range of integration from $10$ to $3000$Hz.  The cross-correlation
analysis with 2-year run of the current LIGO detector pair has
placed an upper limit $\Omega_{\rm GW}< 7.2\times 10^{-6}$
\cite{Abbott:2009ws}, which was performed for the frequency band
$41.5$-$169.25{\rm Hz}$.

\subsection{Accessible parameter space of cosmic string search}

\begin{figure}[p]
\begin{minipage}{1\hsize}
\begin{center}
\subfigure[$p=1$]{
\includegraphics[width=80mm]
{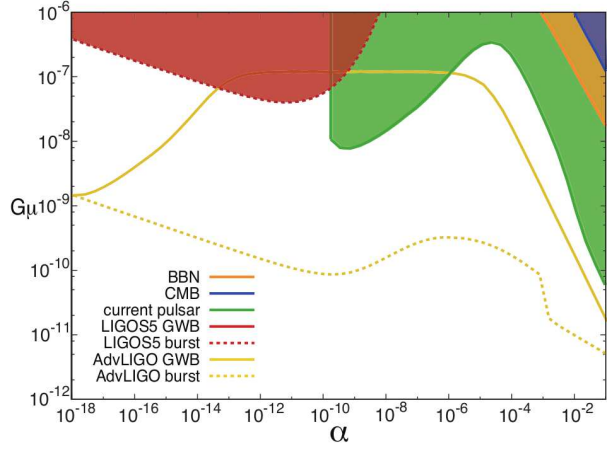}
\label{fig:alpha-Gmu_p=1}
}

\subfigure[$p=10^{-1}$]{
\includegraphics[width=80mm]
{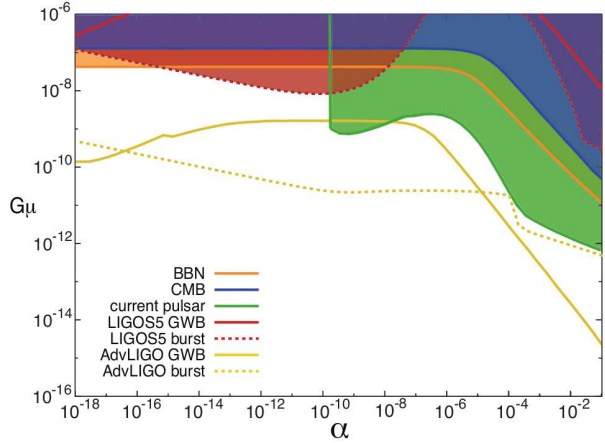}
\label{fig:alpha-Gmu_p=1e-1}
}

\subfigure[$p=10^{-2}$]{
\includegraphics[width=80mm]
{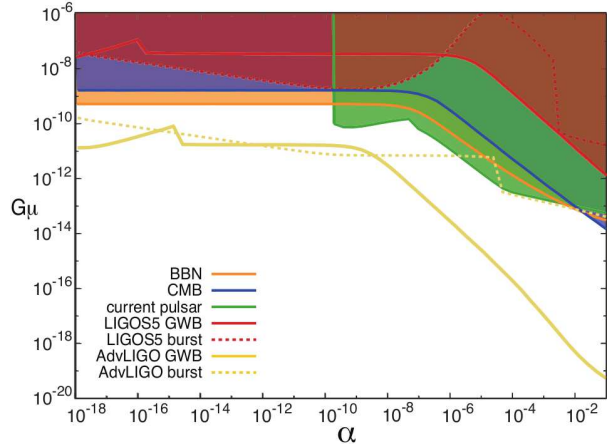}
\label{fig:alpha-Gmu_p=1e-2}
}

\end{center}
\end{minipage}
\caption{The parameter regions excluded by current experiments and
  cosmological constraints, as well as regions which can be probed by
  Advanced LIGO, for different values of $p$.  The colored regions are
  excluded by current experiments.  Advanced LIGO can probe the region
  above the yellow solid line with GW background search and that above
  the yellow dotted line with burst search.  }
\end{figure}

Here, we find the parameter space excluded by current GW experiments
and cosmological constraints, and that accessible by future GW
experiments.  In Fig \ref{fig:alpha-Gmu_p=1},
\ref{fig:alpha-Gmu_p=1e-1}, and \ref{fig:alpha-Gmu_p=1e-2}, we show
the parameter regions which are excluded or can be probed by GW
experiments in the $\alpha-G\mu$ plane for $p=1$, $10^{-1}$, and
$10^{-2}$, respectively.  Current constraints are provided by pulsar
timing and current LIGO experiments.  We also show the cosmological
constraints from CMB and BBN, which comes from the fact that the
energy density of the GW background, which has the same effect as
extra neutrino species on CMB and BBN, must be small at the last
scattering epoch and BBN, so as not to distort the fluctuation of CMB
and change abundance of various nuclei.  For interferometers, we show
both parameter regions accessible by burst and GW background search
with GW detector network consists of Advanced LIGO and other next
generation detectors.  For burst detection, we define the detection
criterion to be whether the bursts whose amplitude is equal to the
detector's best sensitivity come with a rate higher than $1{\rm
  yr}^{-1}$.  Following the discussion in the previous section, we
take the best sensitivity of current LIGO as $fh=1.7\times 10^{-21}$
at $f=150{\rm Hz}$, and the best sensitivity of Advanced LIGO with GW
detector network as $fh=3.4\times 10^{-23}$ at $f=220{\rm Hz}$.  For
the GW background search, we assume it is detectable if the amplitude
is higher than $\Omega_{\rm GW}=7.2\times 10^{-6}$ for LIGO, and
$\Omega_{\rm GW}=3.6 \times 10^{-9}$ for Advanced LIGO.

For the upper limit of $\Omega_{\rm GW}$ from current pulsar timing
experiments, we take $1.9\times 10^{-8}$ at $f=3.2\times 10^{-8} {\rm
  Hz}$ \cite{Jenet:2006sv}.  The CMB provides the constraint that
$\int\Omega_{\rm GW} (f) d (\ln f)$ must be less than $1.4\times
10^{-5}$ at the last scattering \cite{Smith:2006nka}.  The BBN
constraint is $\int\Omega_{\rm GW} (f) d (\ln f) < 1.6\times 10^{-5}$
at the epoch of BBN \cite{Siemens:2006yp,Cyburt:2004yc}.  The lower
limit of the integral is determined by the lowest frequency of the GWs
emitted by largest and youngest loops at the time of CMB and BBN.  The
upper limit is the frequency of GWs emitted by the earliest loops when
they appeared.  We consider that the earliest loops are
formed at the end of the friction domination, when the temperature of
the Universe is $\sim \sqrt{G}\mu$.

The reason why the current pulsar timing experiments constrain only
the large $\alpha$ region is that, for a small value of $\alpha$,
loops cannot generate GWs at such low frequencies accessible by pulsar
timing experiments, which is comparable to $1{\rm yr}^{-1}$.  We see
that although current constraints on the parameter region are rather
severe especially when we consider large $\alpha$, there are still
allowed regions which can be probed by Advanced LIGO with both burst
detection and the GW background.  Choosing parameter values from such
a region, we calculate the Fisher information matrix to study how the
cosmic string parameters can be constrained by future GW experiments
in the following sections.

\subsection{Formalism of Fisher analysis}
The maximum likelihood method is widely used to estimate model
parameters in the analysis of cosmological observations
\cite{Tegmark:1996bz}.  For a given data set, the set of parameters
that is most likely to result in the model prediction are those which
maximize likelihood function ${\cal L}$.  The error in this estimation
can be predicted by calculating the Fisher information matrix, which
is defined as
\begin{equation}
{\cal F}_{lm}\equiv -\frac{\partial^2\ln{\cal L}}{\partial\theta_l\partial\theta_m}.
\label{Fisher}
\end{equation}
Assuming a Gaussian likelihood, the expected error in the parameter
$\theta_l$ is given by 
\begin{equation}
\sigma_{\theta_l}=\sqrt{ ({\cal
    F}^{-1})_{ll}}
\end{equation}
Here, we apply this Fisher matrix formalism to predict how accurately
cosmic string parameters can be constrained in future direct detection
experiments, for both cases of burst and stochastic background
detection.

\subsubsection{Burst detection with a single detector}
If GW bursts from cosmic strings are detected frequently by GW
detectors, we would be able to make a catalogue of the bursts from
cosmic strings.  Let us suppose that we have a large enough number of
samples and the number of observed GW bursts per strain interval $h_i$
to $h_i+dh_i$ is given by
\begin{equation}
N_i=\Phi (h_i) dh_i,
\label{eqPhi}
\end{equation}
where $\Phi (h)\equiv dR/dh\times T$ is predictable from the cosmic
string parameters as presented in Sec \ref{secGW}.  Depending on the
detector sensitivity, the catalogue has a magnitude limit of
$h=h_{\min}$.

Assuming that the number of GW bursts follows a Poisson distribution
\cite{Siemens:2006vk,gr-qc/0312056,arXiv:0710.0497,arXiv:1004.3499},
the probability of observing $k_i$ events in each strain bin is given by
\begin{equation}
p_i=\frac{(N_i)^{k_i}e^{-N_i}}{k_i!}.
\end{equation}
Here, $N_i$ is a function of $\theta_l$, which can be predicted for
given parameters as given in Eq. (\ref{eqPhi}).  The likelihood
function is defined by the total probability of all bins as
\begin{equation}
{\cal L}=\prod_i \frac{(N_i)^{k_i}e^{-N_i}}{k_i!}.
\end{equation}
Substituting the likelihood into Eq. (\ref{Fisher}), we obtain the
Fisher matrix
\begin{eqnarray}
{\cal F}_{lm}&=&-\frac{\partial^2}{\partial\theta_l\partial\theta_m}\left[\sum_i(k_i\ln N_i+N_i-\ln k_i!)\right]\\
&=&-\sum_i\left[k_i\left(-\frac{\partial N_i}{\partial\theta_l}\frac{\partial N_i}{\partial\theta_m}\frac{1}{N_i^2}+\frac{\partial^2 N_i}{\partial\theta_l\partial\theta_m}\frac{1}{N_i}\right)+\frac{\partial^2 N_i}{\partial\theta_l\partial\theta_m}\right]\\
&=&\sum_i \frac{\partial N_i}{\partial\theta_l}\frac{\partial N_i}{\partial\theta_m}\frac{1}{N_i}.
\end{eqnarray}
In the last step, we have used $k_i\rightarrow N_i$ when averaged over
large samples.  Substituting Eq. (\ref{eqPhi}) and rewriting in terms
of integral, the Fisher matrix is given by
\begin{equation}
{\cal F}_{lm}=\int^{\infty}_{h_{\rm min}}\frac{\partial\Phi}{\partial\theta_l}\frac{\partial\Phi}{\partial\theta_m}\frac{1}{\Phi}dh.
\end{equation}

We use bursts detectable by the future GW detector network with $\rho
>4$, which corresponds to the limit of the catalogue being $fh_{\rm
  min}\sim 3.4\times 10^{-23}$ at $f=220{\rm Hz}$, as mentioned in
Sec. \ref{burst_detectability}. \footnote{ Higher signal to noise
  ratio may be required for the use of the Fisher matrix
  approximation, since the distribution of the amplitude $A$ deviates
  from the Gaussian shape because of uncertainties in determining the
  sky location of the burst \cite{Cohen:2010xd}.  Here, however, we
  assume that the sky locations of bursts are determined with a good
  accuracy by using the multiple detector network.  }

\subsubsection{Search for the stochastic GW background}
The analysis of cross correlation is performed in the frequency domain
\cite{Abbott:2003hr,Seto:2005qy}.  Let us consider frequency bins,
each of which has a center frequency $f_i$ and the width $\delta f_i$.
We assume the width is much larger than frequency resolution $\delta
f_i/\Delta f\gg 1$, where $\Delta f\equiv T^{-1}$, so that each bin is
statistically independent.  Describing Eqs. (\ref{Eqmu}) and
(\ref{Eqsigma}) in terms of the discrete Fourier transform, the
cross-correlated signal and its variance are rewritten as
\begin{eqnarray}
\langle \mu\rangle =2\sum_i\frac{3H_0^2}{20\pi^2}\frac{\delta f_i}{\Delta f}f_i^{-3}\gamma_{IJ}(f_i)\Omega_{\rm GW}(f_i)\tilde{Q}(f_i)
\equiv\sum_i\langle \mu_i\rangle ,
\label{signal_i}
\end{eqnarray}
\begin{equation}
\sigma^2=2\sum_i\frac{1}{4}\frac{\delta f_i}{\Delta f}S_{n,I}(f_i) S_{n,J}(f_i)|\tilde{Q}(f_i)|^2\equiv\sum_i\sigma_i^2.
\label{noise_i}
\end{equation}
Assuming Gaussian distribution of the data $\hat{\mu}_i$ around the
mean value $\langle \mu_i\rangle$ with the variance $\sigma_i$ in each
segment, the probability distribution function is given by
\begin{equation}
p_i=\frac{1}{\sqrt{2\pi\sigma_i^2}}\exp\left[-\frac{(\hat{\mu}_i-\langle \mu_i\rangle )^2}{2\sigma_i^2}\right].
\end{equation}
Then, the likelihood function is defined by the total probability for
the whole sample as
\begin{equation}
{\cal L}=\prod_i \frac{1}{\sqrt{2\pi\sigma_i^2}}\exp\left[-\frac{(\hat{\mu}_i-\langle \mu_i\rangle )^2}{2\sigma_i^2}\right].
\end{equation}
Substituting the likelihood into Eq. (\ref{Fisher}), we obtain the
Fisher matrix
\begin{eqnarray}
{\cal F}_{lm}&=&-\frac{\partial^2}{\partial\theta_l\partial\theta_m}\sum_i\left[-\frac{(\hat{\mu}_i-\langle \mu_i\rangle )^2}{2\sigma_i^2}\right]\nonumber\\
&=&-\sum_i\frac{1}{\sigma_i^2}\left[-\frac{\partial\langle\mu_i\rangle}{\partial\theta_l}\frac{\partial\langle\mu_i\rangle}{\partial\theta_m}+(\hat{\mu}_i-\langle \mu_i\rangle )\frac{\partial^2\langle \mu_i\rangle}{\partial\theta_l\partial\theta_m}\right]\nonumber\\
&=&\sum_i \frac{1}{\sigma_i^2}\frac{\partial\langle\mu_i\rangle}{\partial\theta_l}\frac{\partial\langle\mu_i\rangle}{\partial\theta_m}.
\end{eqnarray}
where the second term of the second line vanishes because
$(\hat{\mu_i}-\langle \mu_i\rangle)$ is zero when averaged over.
Substituting $\langle \mu_i\rangle$ and $\sigma_i$ defined in
Eqs. (\ref{signal_i}) and (\ref{noise_i}), the Fisher matrix is given
as
\begin{equation}
{\cal F}_{lm,IJ}=\left (\frac{3H_0^2}{10\pi^2}\right)^2 2T
\int^{\infty}_{0}df\frac{|\gamma_{IJ} (f)|^2\partial_{\theta_l}
\Omega_{\rm GW} (f)\partial_{\theta_m}\Omega_{\rm
GW} (f)}{f^6S_{n,I} (f)S_{n,J} (f)},
\end{equation}
For multiple detectors, the Fisher matrix can be calculated with 
\begin{equation}
{\cal F}_{lm}=\left (\frac{3H_0^2}{10\pi^2}\right)^2 2T\sum^N_{I=1}\sum^N_{J<I}
\int^{\infty}_{0}df\frac{|\gamma_{IJ} (f)|^2\partial_{\theta_l}
\Omega_{\rm GW} (f)\partial_{\theta_m}\Omega_{\rm
GW} (f)}{f^6S_{n,I} (f)S_{n,J} (f)}.
\end{equation}

\subsection{
Constraints on cosmic string parameters}
\begin{figure}
\begin{minipage}{1\hsize}
 \begin{center}
 \subfigure[$p-\alpha$]{
  \includegraphics[width=0.45\textwidth]{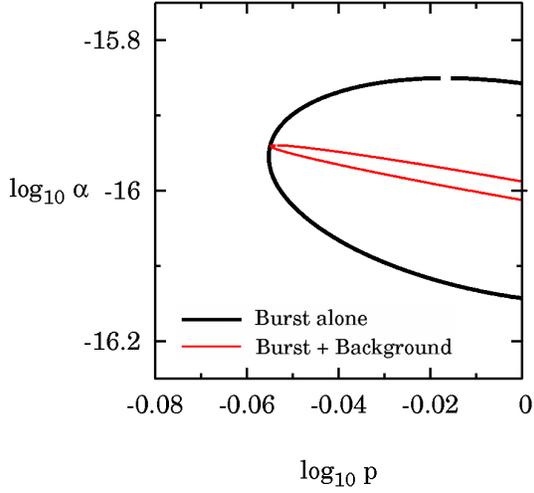}
  }
  \subfigure[$\alpha-G\mu$]{
  \includegraphics[width=0.45\textwidth]{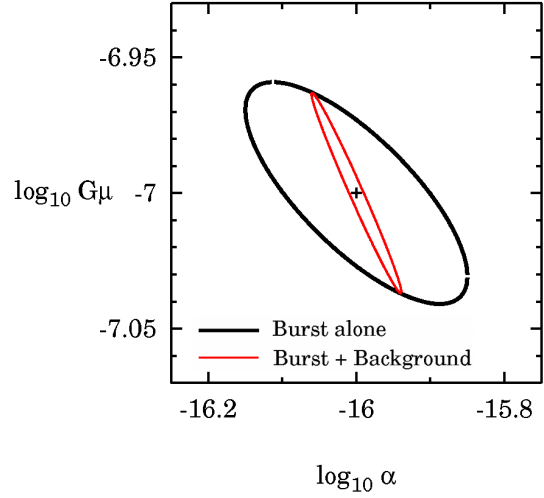}
  }
  \subfigure[$p-G\mu$]{
  \includegraphics[width=0.45\textwidth]{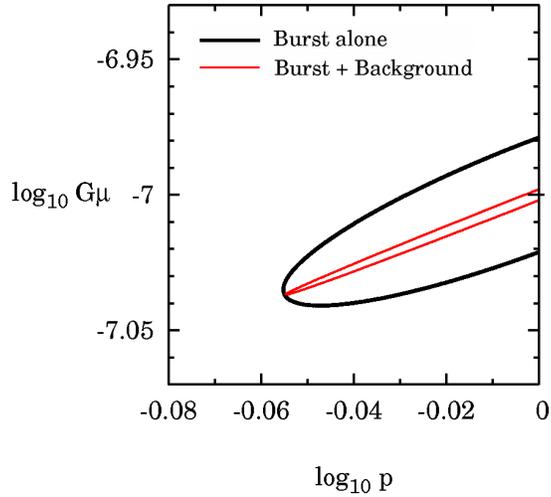}
  }
 \end{center}
 \end{minipage}
 \caption{\label{contour} Marginalized $2\sigma$ constraints on cosmic
   string parameters in $p-\alpha$, $\alpha-G\mu$, and $p-G\mu$
   planes, respectively.  The fiducial parameters are taken to be
   $G\mu=10^{-7}$, $\alpha=10^{-16}$, $p=1$.  The solid black line
   represents the constraints from the burst detection alone,
   and the red line represents the combined constraints from the burst
   detection and the GW background measurement.}
\end{figure}

Using the Fisher matrix formalism presented above, we forecast
constraints on cosmic string parameters from future direct detection
experiments.  In Fig. \ref{contour}, we present an example of the
expected future constraints in the case where the parameters are
$G\mu=10^{-7}$, $\alpha=10^{-16}$, $p=1$.  Each ellipse represents the
$2\sigma$ error contours expected from 3 years of observation with
future ground-based GW network (Advanced LIGO, KAGRA and VIRGO with
the sensitivity given in Eq.  (\ref{Adv-LIGO})).  In this setup,
$1.94\times 10^5$ bursts are detected with $\rho>4$ by multiple
detectors in 3-year observation, and the GW background is detected
with $\rho\sim 161$ by correlating outputs of all the detectors.  We
clearly see the constraints from burst detection (black line) are
tighten when it is combined with that from the GW background (red
line).

The reason why we can obtain the stronger constraints by combining
burst detection and GW background measurements is that the constraints
from the two measurements have different parameter degeneracies.  In
the case of the burst detection, the direction of the parameter
degeneracy is determined by the parameter dependence of the rate
$dR/d\ln h$, which is presented in Appendix \ref{app1}.  The parameter
set used here corresponds to the case (iii), i.e. Eq.
(\ref{dR_dh:case3}), with $fh_{3,3}\simeq 4.7\times 10^{-25}$.  Since
GWs larger than $h_{3,3}$ are observed via burst detection, the
direction of the parameter degeneracy is $\propto (G\mu)^2
\alpha^{1/3} p^{-1}$.  In the case of the GW background, the direction
of the parameter degeneracy is determined by the parameter dependence
of $\Omega_{\rm GW}$, which is investigated in Appendix \ref{app2}.
In this parameter set (the case (iii)), the GW background is given by
(\ref{OmegaGW3}), and the second term dominates the first term if it
is evaluated at the frequency of Advanced LIGO.  Thus, the constraints
from the GW background has the parameter degeneracy of $\propto G\mu
\alpha^{-1/3} p^{-1}$.  The constraints from the GW background have
strong parameter degeneracies since the observable is basically only
$\Omega_{\rm GW}$ at $f=220$Hz.  By combining it with the constraints
from bursts, the degeneracies are broken. \footnote{ Note that, the
  direction of the degeneracy seen in the figures does not directly
  correspond to the parameter dependence described here, since the
  shown constraints are marginalized over the other parameter.  }
  
Here, we would like to mention when these bursts and the background
detectable by Advanced LIGO are emitted in the case of the parameter
set we take above.  As mentioned in Appendix \ref{app1}, bursts of
given frequency $f$ and amplitude $h$ are mainly emitted at redshift
which satisfies $l (f,h,z) \sim \alpha t (z)$ for $\alpha < \Gamma
G\mu$, so bursts detectable by Advanced LIGO are generated at $z\ll
1$.  To be more specific, GWs of amplitude equal to the sensitivity of
Advanced LIGO, $fh\simeq 3.4\times 10^{-23}$, and of frequency
$f=220{\rm Hz}$ come from $z\sim 3.6\times 10^{-2}$, and those
detected once per year by Advanced LIGO, which have $fh\sim 10^{-21}$
and $f=220{\rm Hz}$, come from $z\sim 10^{-3}$.  The main contribution
on the GW background comes from GWs emitted at redshift which
satisfies $f (1+z)\sim \alpha t (z)$ in the radiation-dominated epoch
and those emitted at $z\sim 1$.  At the frequency of Advanced LIGO,
the contribution from the latter is larger.

\section{Conclusion}
Future GW experiments can be a unique and useful tool to test the
existence of cosmic strings.  If GWs from cosmic strings are detected,
they could provide important constraints on cosmic string parameters.
In this paper, we have studied the potential of upcoming ground-based
GW experiments to search for signals from cosmic strings and estimated
the power to place constraints on cosmic string parameters.

The key point of this paper is that GW experiments can search for
cosmic string signals both in the way of burst detection and GW
background measurements.  First, we find the parameter region where
GWs from cosmic strings are detectable by future experiments
considering the both cases.  Furthermore, we investigate constraints
on cosmic string parameters obtainable from direct detection
measurements of the GW bursts and the GW background, and found that
their information is complementary from each other.  Thus, if both the
GW bursts and the GW background are detected, we can tighten the
constraints on cosmic string parameters by combining data from the two
different measurements.  Although we demonstrate only the case of
ground-based experiments, this is also the case in future satellite
experiments like BBO and DECIGO which is designed to search for both
GW bursts and GW backgrounds.

One thing we must note is that there would be other sources of GW
bursts and GW backgrounds.  There are surely many astrophysical
candidates which generate GW bursts.  However, since GW bursts from
cosmic strings have a characteristic frequency dependence, cosmic
string bursts could be distinguished from those from other sources.
In the case of the GW background, although there is no certain source,
some models can predict a GW background around the frequency band of
LIGO sensitivity.  We could also use information on the frequency
dependence of the spectrum, but it may be difficult to identify
whether the detected GW background originates from cosmic strings or
other models.  However, although it depends on the values of the
cosmic string parameters, we may be able to use other observations
which explore different frequency ranges of GWs such as CMB $B$-mode
measurements
\cite{Seljak:1997ii,Benabed:2000tr,Seljak:2006hi,Bevis:2007qz,Pogosian:2007gi,Kawasaki:2010iy,Bevis:2010gj,Mukherjee:2010ve}
and pulsar timing experiments \cite{Jenet:2006sv,Pshirkov:2009vb} to
verify the GW background from cosmic string, which will be studied in
our future work.

\appendix

\section*{Appendix}

\section{Parametric dependence of the burst rate }
\label{app1}  
In this appendix, we roughly estimate the dependence of the burst rate
on cosmic string parameters.  Cosmic strings expiring at each epoch
give dominant contribution to GWs, because they are the most
numerous.  So, roughly speaking, we only need to estimate their
contribution to the burst rate, as shown formally below.  Since we are
interested in high frequency GWs detectable by ground-based GW
experiments, we consider bursts which satisfy $f>\left (\min
  \{\alpha,\Gamma G\mu\} \right)^{-1} t^{-1} (z_{eq})
(1+z_{eq})^{-1}$, which means that expiring loops start to contribute
to relevant frequencies in the radiation-dominated era and their
contribution continues until today.

The rate $dR/d\ln h$ is derived by integrating Eq. (\ref{dR_dzdh}) in
terms of $z$.  Using Eqs. (\ref{thetam}), (\ref{l_of_hzf}), and
(\ref{t_i}), we find that in the case where $\alpha\gg \Gamma G\mu$,
so loops are long-lived, the dominant contribution to the integration
of Eq. (\ref{dR_dzdh}) comes from the redshift around $z_m$ which
satisfies
\beq
l (f,h,z_m) = \Gamma G\mu t (z_m), \label{z_m}
\eeq
for large $h$.  The size of loops which emit GW bursts of frequency
$f$ and amplitude $h$ depends on when they are emitted and it is
denoted by $l (f,h,z)$.  Equation (\ref{z_m}) means that, for a given
frequency and amplitude, GWs are mostly generated by loops expiring at
redshift $z_m$.  Therefore, we can roughly estimate the rate by
$dR/dh\sim z_m (f,h)\times \frac{dR}{dhdz} (f,h,z_m (f,h))$,
neglecting accuracy of $\mathcal{O} (1)$ numerical factors.
\footnote{ We also neglect the dependence on $\Gamma$ in the following
  estimation, since it is expected to be canceled by numerical factors
  which determine $\Gamma$, for example, that in Eq. (\ref{strain}).
} Note that for too small $h$, there is no solution which satisfies
both Eq. (\ref{z_m}) and $f (1+z_m)>l^{-1} (f,h,z_m)$, which
corresponds to the lower frequency cutoff of bursts.  In this case,
the main contribution to the $z$ integral of Eq. (\ref{dR_dzdh}) comes
from $z_m$ which satisfies $f (1+z_m)= l^{-1} (f,h,z_m)$.

In the case of short-lived loops, $\alpha \ll \Gamma G\mu$, $z_m$ is
given by $l (f,h,z_m) = \alpha t (z_m)$.  In this case, if we consider
small $h$, any value of $z$ does not satisfy $l (f,h,z) = \alpha t
(z)$, so $dR/d\ln h$ is strongly suppressed.

The rate of GW burst depends on when bursts are mainly emitted and
when loops which emit such bursts are
formed.  We summarize the results below:\\

(i)  $\alpha>\Gamma G\mu \left (\frac{\Omega_m}{\Omega_r}\right)^{3/2}$

(Here, $\Omega_r$ is the current energy density of radiation divided by the critical density.)\\

In this case, loops are so long-lived that those formed in the
matter-dominated era does not decay before today.  In other words,
loops which have decayed by today are formed in the
radiation-dominated era.

The burst rate is given as follows.  \small

\beq
\frac{dR}{d\ln h} (f,h)  \sim 
\begin{cases}
&  (G\mu)^{1/2}\alpha^{1/2} \gamma_r^{-2}\left (\frac{\Omega_m}{\Omega_r}\right)^{-1/4} f^{3/2}t_0h^{-1/2}  \quad ;h< \left (\frac{\Omega_m}{\Omega_r}\right)^{-1/2} f^{-3}t_0^{-2}\equiv h_{1,1}\\
& (G\mu)^{1/2}\alpha^{1/2} \gamma_r^{-2} \left (\frac{\Omega_m}{\Omega_r}\right)^{-11/10} f^{-18/5} t_0^{-12/5} h^{-11/5}   \\
&\quad  ;h_{1,1}<h< (G\mu)^{5/3} f^{-4/3} t_0^{-1/3} \left (\frac{\Omega_m}{\Omega_r}\right)^{-4/3}\equiv h_{1,2} \\
&  (G\mu)^{1/16}\alpha^{1/2} \gamma_r^{-2} \left (\frac{\Omega_m}{\Omega_r}\right)^{-3/4} f^{-13/4} t_0^{-37/16} h^{-31/16} \\
&\quad  ;h_{1,2}<h< (G\mu)^{5/3}f^{-4/3}t_0^{-1/3} \equiv h_{1,3}\\
&  (G\mu)^{11/6}\alpha^{1/2} \gamma_r^{-2} \left (\frac{\Omega_m}{\Omega_r}\right)^{-3/4} f^{-14/3} t_0^{-8/3} h^{-2}
 \quad ;h>h_{1,3}
\end{cases},
\label{dR_dh:case1}
\eeq
\normalsize
where $t_0$ is the present age of the Universe.

Bursts whose amplitudes are $h_{1,1}<h<h_{1,2}$, $h_{1,2}<h<h_{1,3}$,
and $h>h_{1,3}$ correspond to those emitted in the radiation-dominated
era, in the matter-dominated era, and at redshift smaller than $1$,
respectively.  If $h<h_{1,1}$, any $z_m$ does not satisfy $f
(1+z_m)\sim l^{-1} (f,h,z_m)$.  Therefore, bursts which have amplitude
of $h<h_{1,1}$ are emitted in the radiation-dominated era not by loops
expiring but by loops which still have their lifetime.

Increase of $G\mu$ enhances the amplitude of each burst.  Since bursts
of a given amplitude can come from earlier epochs and more distant
points for larger $G\mu$, their number increases.  This is why the
rate is proportional to a positive power of $G\mu$.  However, there is
also an effect which decreases the rate with increasing $G\mu$, that
is, reduction of the loop lifetime and the density of expiring loops
due to the enhancement of the efficiency of the GW emission.  Note
that loops formed earlier are more abundant than those formed later
despite more dilution by the cosmic expansion, since the density at
their formation is larger for older loops.  The total power of $G\mu$
is determined by these effects.  As $\alpha$ increases, the elongation
of loop lifetime enhances the rate but the initial number of loops
decreases.  In this case, the former is more efficient and the power
of $\alpha$ in Eq.  (\ref{dR_dh:case1})
becomes positive.\\

 (ii) $\Gamma G\mu<\alpha<\Gamma G\mu \left (\frac{\Omega_m}{\Omega_r}\right)^{3/2}$\\

 In this case, loops which expire at small redshift are formed in the
 matter-dominated era.  The burst rate is given by

\small

\beq
\frac{dR}{d\ln h} (f,h) \sim 
\begin{cases}
&  (G\mu)^{1/2}\alpha^{1/2} \gamma_r^{-2}\left (\frac{\Omega_m}{\Omega_r}\right)^{-1/4} f^{3/2}t_0h^{-1/2}  \quad ;h< \left (\frac{\Omega_m}{\Omega_r}\right)^{-1/2} f^{-3}t_0^{-2}\equiv h_{2,1}\\
&  (G\mu)^{1/2}\alpha^{1/2} \gamma_r^{-2} \left (\frac{\Omega_m}{\Omega_r}\right)^{-11/10} f^{-18/5} t_0^{-12/5} h^{-11/5} \\
& \quad  ;h_{2,1}<h<  (G\mu)^{5/3} f^{-4/3} t_0^{-1/3} \left (\frac{\Omega_m}{\Omega_r}\right)^{-4/3} \equiv h_{2,2} \\
&  (G\mu)^{1/16}\alpha^{1/2} \gamma_r^{-2} \left (\frac{\Omega_m}{\Omega_r}\right)^{-3/4} f^{-13/4} t_0^{-37/16} h^{-31/16} \\
& \quad  ;h_{2,1}<h< (G\mu)^{7/9}\alpha^{8/9}f^{-4/3}t_0^{-1/3} \left (\frac{\Omega_m}{\Omega_r}\right)^{-4/3} \equiv h_{2,3}\\
& (G\mu)^{-3/8}\gamma_m^{-2} f^{-5/2} t_0^{-17/8} h^{-11/8} \\
& \quad ;h_{2,2}<h< (G\mu)^{5/3}f^{-4/3}t_0^{-1/3}\equiv h_{2,4} \\
&  (G\mu)^{7/3} \gamma_m^{-2} f^{-14/3} t_0^{-8/3} h^{-3}  \quad ;h>h_{2,4}
\end{cases}.
\label{dR_dh:case2}
\eeq

\normalsize

Bursts whose amplitudes are $h_{2,1}<h<h_{2,2}$ and
$h_{2,2}<h<h_{2,3}$ correspond to those emitted in the
radiation-dominated era and in the matter-dominated era by loops
formed in the radiation-dominated era.  On the other hand, bursts in
the range of $h_{2,3}<h<h_{2,4}$ and $h>h_{2,4}$ are emitted at
redshift $1<z<z_{eq}$ and $z<1$ by loops formed after the
matter-radiation equality.  These are mainly emitted by expiring
loops.  Bursts of $h<h_{2,1}$, on the
other hand, are emitted by loops which do not expire soon.\\

 (iii) $\alpha<\Gamma G\mu$\\

 In this case, loops are short-lived.  The rate is given by

\small

\beq
\frac{dR}{d\ln h} (f,h) \sim
\begin{cases}
&  (G\mu)^{6/5}\alpha^{-1/5} \gamma_r^{-2} \left (\frac{\Omega_m}{\Omega_r}\right)^{-11/10} f^{-18/5} t_0^{-12/5} h^{-11/5}\\
& \qquad  ;G\mu\alpha^{-1}f^{-3}t_0^{-2}\left (\frac{\Omega_m}{\Omega_r}\right)^{-1/2}\equiv h_{3,1}<h<G\mu\alpha^{2/3} f^{-4/3} t_0^{-1/3} \left (\frac{\Omega_m}{\Omega_r}\right)^{-4/3} \equiv h_{3,2} \\
&  (G\mu)^{3/8}\alpha^{-3/4} \gamma_m^{-2} f^{-5/2} t_0^{-17/8} h^{-11/8} \\
& \qquad  ;h_{3,2}<h<G\mu\alpha^{2/3}f^{-4/3}t_0^{-1/3}\equiv h_{3,3} \\
&  (G\mu)^{2}\alpha^{1/3} \gamma_m^{-2} f^{-14/3} t_0^{-8/3} h^{-3}  \qquad ;h>h_{3,3}
\end{cases}.
\label{dR_dh:case3}
\eeq

\normalsize

Bursts in the range of $h_{3,1}<h<h_{3,2}$, $h_{3,2}<h<h_{3,3}$, and
$h>h_{3,3}$ are emitted in the radiation-dominated era, in the
matter-dominated era, and at $z\ll 1$, respectively.  The burst rate
for $h<h_{3,1}$ is suppressed for the same reason as $h<h_{1,1}$ and
$h<h_{2,1}$.  In these cases, elongation or shortening of loop
lifetime do not affect the rate and the rate depends only on
parameters through time of burst emission and initial number of loops.

\section{Parametric dependence of the spectrum of the stochastic GW background} 
\label{app2}  

In this appendix, we roughly describe the amplitude of $\Omega_{\rm
  GW}$ in terms of cosmic string parameters.  In the same way as the
previous section, we consider only high frequency GWs. 

 (i)  $\alpha>\Gamma G\mu \left (\frac{\Omega_m}{\Omega_r}\right)^{3/2}$\\

 In this cases, we find through Eq.  (\ref{dR_dh:case1}) that the
 dominant contribution to the integral of $h^2\frac{dR}{dh}$ in Eq.
 (\ref{OmegaGW}) comes from 
$h\sim h_{1,1}$ or $h\sim h_{1,3}$.
 Approximating that the contribution to $\int d\ln h h^3
 \frac{dR}{dh}$ only comes from these GWs simply as
 $h_m^3\frac{dR}{dh} (h_m)$, where $h_m$ is $h_{1,1}$ or $h_{1,3}$, we
 get
\beq
\Omega_{\rm GW}\sim  (G\mu)^{1/2} \alpha^{1/2} \gamma_r^{-2} \frac{\Omega_r}{\Omega_m} +  (G\mu)^{1/6}\alpha^{1/2}\gamma_r^{-2}\left ( \frac{\Omega_m}{\Omega_r}\right)^{-3/4}t_0^{-1/3}f^{-1/3}. \label{OmegaGW1}
\eeq
The first term, the contribution from $h\sim h_{1,1}$, represents the
GWs emitted when loops decay in the radiation-dominated era, and the
second one, the contribution from $h\sim h_{1,3}$, corresponds to GWs
emitted at $z\sim 1$ by loops which expired recently or are decaying
today.

We can estimate that the first term is equal to the energy of loops
expiring at time $t$ (redshift $z$) which satisfies $\Gamma G\mu t\sim
1/f (1+z)$ as
\begin{align}
\Omega_{\rm GW} &\sim \frac{\rho_{loop} (t)|_{l\sim \Gamma G\mu t}}{\rho_{tot} (t)}\frac{\Omega_r}{\Omega_m} \nonumber \\
&\sim \frac{\rho_{loop} (t_i)|_{l \sim \alpha t_i}}{\rho_{tot} (t_i)} \frac{1+z_i}{1+z}\frac{\Omega_r}{\Omega_m} \nonumber \\
&\sim G\mu \gamma_r^{-2}\left (\frac{\alpha}{\Gamma G\mu}\right)^{1/2}\frac{\Omega_r}{\Omega_m} \nonumber \\
&\sim   (G\mu)^{1/2}\alpha^{1/2} \gamma_r^{-2} \frac{\Omega_r}{\Omega_m}.
\label{OmegaGW1_1}
\end{align}
Here, $t_i$ and $z_i$ are the time and redshift of the loop formation
respectively, which is related to the time of GW emission as $t_i\sim
\frac{\Gamma G\mu}{\alpha}t$, and the factor
$\frac{\Omega_r}{\Omega_m}$ denote the dilution of GWs from the
matter-radiation equality to now.

The second term can be estimated as follows,
\beq
\Omega_{\rm GW}  \sim \frac{1}{\gamma_r^2\alpha t_i^3} (1+z_i)^3 \times\frac{d\dot{E}}{df} (f)f\times t_0\times \frac{1}{\rho_{cr}} \sim  (G\mu)^{1/6}\alpha^{1/2}\gamma_r^{-2}\left (\frac{\Omega_m}{\Omega_r}\right)^{-3/4}t_0^{-1/3}f^{-1/3}, \label{OmegaGW1_2}
\eeq
where $t_i\sim \frac{\Gamma G\mu}{\alpha}t_0$ and $z_i$ are the time
and the redshift when loops expiring today are formed.  Note that the
number of GWs emitted recently is so small that it might not be
regarded as a part of the background.  The condition $\frac{dR}{d\ln
  h} (h=h_{1,3}) > f$ leads to
\beq
 (G\mu)^{-19/6}\alpha^{1/2}\gamma_r^{-2}\left (\frac{\Omega_m}{\Omega_r}\right)^{-3/4} (ft_0)^{-5/3}>1,
\eeq
which can be satisfied by small $G\mu$, 
for example $G\mu \lesssim 10^{-11}$, for $f\sim 100{\rm Hz},
\alpha=0.1, p=1$.\\

 (ii) $\Gamma G\mu<\alpha<\Gamma G\mu \left (\frac{\Omega_m}{\Omega_r}\right)^{3/2}$\\

One can get $\Omega_{\rm GW}$ by estimating $\int dh h^2
\frac{dR}{dh}$ in the same way as demonstrated above.  The result is
\beq
\Omega_{\rm GW}\sim  (G\mu)^{1/2} \alpha^{1/2} \gamma_r^{-2} \frac{\Omega_r}{\Omega_m} +  (G\mu)^{2/3}\gamma_m^{-2}t_0^{-1/3}f^{-1/3}. \label{OmegaGW2}
\eeq
Again, the first term corresponds to the GWs emitted by loops decayed
in the radiation-dominated era and the second one is the contribution
from loops expiring today.  The first term is identical to the one in
Eq. (\ref{OmegaGW1}).  On the other hand, the second one is different
because loops expiring today are formed after matter-radiation
equality in this case.

The condition to include the second term in $\Omega_{\rm GW}$ is
\beq
 (G\mu)^{-8/3}\gamma_m^{-2} (ft_0)^{-5/3}>1.
\eeq
For $f\sim 100{\rm Hz}$ and $p=1$, this reads $G\mu\lesssim 10^{-12}$.\\

 (iii) $\alpha<\Gamma G\mu$\\

In this case, we get 
\beq
\Omega_{\rm GW}\sim G\mu \gamma_r^{-2} \frac{\Omega_r}{\Omega_m} + G\mu\gamma_m^{-2}\alpha^{-1/3}t_0^{-1/3}f^{-1/3}. \label{OmegaGW3}
\eeq
Again, the first term is the contribution from GWs emitted in the
radiation-dominated era and the second one represents GWs emitted
recently.  The first term coincides with the ratio of the energy of
the infinite string network to the total energy, $\sim
G\mu\gamma^{-2}$, except the dilution factor
$\frac{\Omega_r}{\Omega_m}$, because, in this case, loops released
from infinite strings immediately expire by radiating GWs.  The second
term also consists of $G\mu\gamma^{-2}$ and the factor $ (f\alpha
t_0)^{-1/3}$, which corresponds to the tilt of the spectrum of emitted
GWs.

The condition to include the second term in $\Omega_{\rm GW}$ is
\beq
 (G\mu)^{-1}\alpha^{-5/3}\gamma_m^{-2} (ft_0)^{-5/3}>1,
\eeq
which is relatively easy to satisfy for small $\alpha$.  For example,
this reads $G\mu \lesssim 10^{-6}$ for $f\sim 100{\rm Hz}$,
$\alpha=10^{-16}$ and $p=1$.

\section*{Acknowledgment}

We appreciate Masahiro Kawasaki for useful discussions.  S. K. would
like to thank Naoki Seto, Takahiro Tanaka, Atsushi Nishizawa, Naoki
Yasuda, Tsutomu Takeuchi, Masaomi Tanaka, and Kohki Konishi for their
helpful advice and discussion.  This work is supported by Grant-in-Aid
for Scientific research from the Ministry of Education, Science,
Sports, and Culture (MEXT), Japan, No.\ 23.10290 (K.M.) and No.\
23740179 (K.T.).  K.M. and T.S. would like to thank the Japan Society
for the Promotion of Science for financial support.

{}

\end{document}